\documentclass[rnote]{aa}
\usepackage{txfonts}
\usepackage{epsf}

\begin{document}

\title{Time lag between prompt optical emission and $\gamma$-rays in GRBs}
\author{Su Min Tang \inst{1}
\and Shuang Nan Zhang \inst{1,2,3,4}}


\offprints{Su Min Tang, \\
\email{tangsm99@mails.tsinghua.edu.cn}}

\institute{Department of Physics and Center for Astrophysics,
Tsinghua University, Beijing 100084, China
\and Department of
Physics, University of Alabama in Huntsville, Optics Building
201C, Huntsville, AL 35899
\and Space Science Laboratory, NASA
Marshall Space Flight Center, SD50, Huntsville, AL 35812
\and
Institute of High Energy Physics, Chinese Academy of Sciences,
P.O. Box 918-3, Beijing 100039, China}

\date{Received <date> / Accepted <date>}
\abstract
{
}
{
The prompt optical emission contemporaneous with the $\gamma$-rays
from $\gamma$-ray bursts (GRBs) carries important information on
the central engine and explosion mechanism. We study the time lag
between prompt optical emission and $\gamma$-rays in GRB 990123
and GRB 041219a, which are the only two GRBs had been detected at
optical wavelengths during the ascending burst phase.}
{
Assuming profiles of prompt optical light curves are the same as the prompt
$\gamma$-rays, we simulate optical light curves with different time lags and compare
them with the observed optical flux. Then the best fit time lag and its error are
determined by chi-squared values.}
{
We find that time lags between prompt optical emission and
$\gamma$-rays in GRB host galaxy rest-frames are consistent in the
two GRBs, which is $5\sim7$ s for GRB 990123 and $1\sim5$ s for GRB
041219a. This result is consistent with a common origin of prompt
optical and $\gamma$-ray emissions in the two GRBs. Based on
synchrotron cooling model, we also derive the parameters for the two
GRBs.}
{
}

\maketitle

\section{Introduction}
The prompt optical emission contemporaneous with the $\gamma$-rays
from $\gamma$-ray bursts (GRBs) carries crucial information on the
central engine and explosion mechanism. Internal shock models and
external shock models have been proposed to explain the prompt
optical emission (see e.g. Meszaros \& Rees 1999; Liang et al.
1999). However, the origin of the prompt optical emission remains
an open issue.

Prompt optical emission have been found for several GRBs during the brief durations of
bursts (e.g. Akerlof et al. 1999; Vestrand et al. 2005; Rykoff et al. 2005; Klotz et
al. 2006). For two of them, i.e. GRB 990123 (Akerlof et al. 1999; Kippen et al. 1999)
and GRB 041219a (Vestrand et al. 2005; Barthelmy, S., et al. 2004), early-time optical
observations were carried out during their bursting phase, including both ascending and
descending parts. Hence, it is possible to study the time lag between prompt optical
emission and $\gamma$-rays in these two GRBs, which could provide important clues to
the origin of prompt optical emission.

\section{Data Analysis}
As shown in Figure~1, in each GRB, there are only three points with a positive optical
detection (crosses) during the burst phase and the profile of optical light curve
remains unknown. To avoid redundant parameters, we simply assume that the prompt
optical light curve profile is exactly the same as the $\gamma$-rays, but with a time
lag to be determined. Then we simulate optical light curves with different time lags
and compare them with the observed optical flux. The redshift of GRB 990123 is 1.6
(Andersen et al. 1999; Kulkarni et al. 1999). The redshift of GRB 041219a is unknown.
Barkov $\&$ Bisnovatyi-Kogan (2005) gave a redshift upper limit as $z\leq0.12$ based on
a model of a dust reradiation of IR afterglow in the envelope surrounding the GRB
source. In this note, we assume $z=0.1$ for GRB 041219a.

We consider the following four models for time lag between optical
and gamma-ray emissions:
\begin{enumerate}
  \item No time lag between optical and gamma-ray emissions in GRB 041219a;
  \item No time lag between optical and gamma-ray emissions in both GRB 990123 and GRB 041219a;
  \item Time lags between optical and gamma-ray emissions in GRB
  990123 and GRB 041219a are independent.
  \item Time lags between optical and gamma-ray emissions in GRB
  990123 and GRB 041219a are the same in their host galaxy rest-frames;
\end{enumerate}

Results of $\chi^2$ tests for the four models are given in Table~1.
For model 3, $\chi^2$ values of different time lags are shown in
Figure~2, and simulated light curves with corresponding lags of
minimum $\chi^2$ values are shown in Figure~1. $\chi^2$ values of
different time lags for model 4 are shown in Figure~3.

As shown in Table~1, model 2 is rejected. Model 4 and 3 are better than model 1,
indicating that a common time lag between prompt optical emission and $\gamma$-rays in
both GRBs is quite possible. In model 3 (Figure~2), for GRB 990123, the best-fitting
time lag is 12.2 s, and there is another local minimum of $\chi^2$ values around 18.2
s. The 12 s lag corresponds to moving the second pulse to the second optical point,
while 18 s corresponds to moving the first pulse to the second optical point. Therefore
the host galaxy rest-frame time lag should be $4.6\sim7.2$ s. For GRB 041219a, the
best-fitting time lag is $3.0^{+2.5}_{-2.3}$ s with $1\sigma$ error, and the host
galaxy rest-frame time lag will be $0.6\sim5.0$ s. The $\chi^2$ value is very sensitive
to the time lag in GRB 990123, thus in model 4 (Figure~3), the best-fitting lag value
is dominated by the contribution of GRB 990123.

\section{Discussion and Conclusions}

It was reported in some previous studies that the behaviors and origins of the prompt
optical emission in GRB 990123 and GRB 041219a are totally different: In GRB 990123,
the optical emission was uncorrelated with the prompt gamma-rays, suggesting that the
optical emission was generated by a reverse external shock arising from the ejecta's
collision with surrounding material; while in GRB 041219a, the optical emission was
correlated with the prompt gamma-rays, indicating optical emission was generated by
internal shocks driven into the burst ejecta by variations of the inner engine
(Vestrand et al. 2005). However, our results show that time correlation between prompt
optical and gamma-ray emission are quite consistent in both GRBs. If GRB 041219a is a
low-redshift object, rest-frame time lags of prompt optical emission behind
$\gamma$-rays in the two GRBs is similar, indicating a common mechanism to produce the
prompt optical emission in the two GRBs.

If both prompt optical and $\gamma$-ray emissions come from cooling $e^{\pm}$ pairs in
the moving ejecta with a Lorentz factor $\Gamma$ through synchrotron radiation,
assuming velocities of $e^{\pm}$ are isotropic, the frequency of emitted photon is
around the peak frequency of synchrotron photons, which in the observer's frame is
\begin{equation}
\nu \approx \nu_m \approx 2\times10^6  \gamma^2B\Gamma/(1+z)\
\mathrm{Hz},
\end{equation}
where $\gamma$ is the Lorentz factor of $e^{\pm}$ and $B$ is the
magnetic field strength in Gauss.

The power radiated per one electron or positron in the observation
frame is
\begin{equation}
P=1.1\times10^{-15}\gamma^2B^2 \Gamma^2/(1+z)^2 \ \mathrm{erg/s}.
\end{equation}

The time lag between prompt optical emission and $\gamma$-rays
should be the typical lifetime of $e^{\pm}$ which emit optical
photons
\begin{equation}
t_{lag} \approx \frac{E}{P}\approx \frac{3.3\times10^8 (1+z)}{\gamma
B^2 \Gamma} \ \mathrm{s},
\end{equation}

Assuming $\Gamma=300$, which is a typical value in GRBs (see e.g. Liang et al. 1999),
we can derive $B$, $\gamma$ and the total $e^{\pm}$ number $N_e$ based on synchrotron
cooling, as listed in Table~2.  Further assuming an emission radius $R=2\times10^{16}$
cm (see e.g. Liang et al. 1999; Li et al. 2003), we can derive the synchrotron
self-absorption frequency $\nu_a$ (Li et al. 2004) and the $e^{\pm}$ column density
$\Sigma$ in the observer's frame
\begin{equation}
\nu_a \approx 1\times 10^{16} L^{2/7}_{50} \Gamma^{3/7}_{300}
R^{-4/7}_{14} B^{1/7}_4/(1+z) \ \mathrm{Hz},
\end{equation}
and
\begin{equation}
\Sigma= \frac{N_e} {\pi (R /\Gamma)^2} \ \mathrm{cm^{-2}}.
\end{equation}

As shown in Table~2, the self-absorption frequencies are less than
the observed frequencies, thus prompt emissions can be observed.
When absorption is negligible, the optical to $\gamma$-ray flux
ratio from synchrotron cooling with constant $\Gamma$ and $B$ should
be $F_{5000\ \AA}/F_{100\ keV}=\nu_{5000\ \AA}/\nu_{100\
keV}=2.5\times10^{-5}$. This is consistent with the observed value
of GRB 041219a, which is $1.2\times10^{-5}$ when there is zero lag
between prompt optical emission and $\gamma$-rays (Vestrand et al.
2005) and about $1.4\times10^{-5}$ for our best-fitting model.
However, in GRB 990123, the peak optical/peak $\gamma$-ray flux
ratio is $3.3 \times 10^{-4}$ (Kulkarni et al. 1999; Akerlof et al.
1999) and the flux ratio changes to be $7.1\times10^{-4}$ for our
best-fitting model when the time lag is 12.2 s, which is one
magnitude larger than the synchrotron-cooling model predicted value,
indicating that such a simplified non-absorption synchrotron cooling
model alone could not account for all the observed properties of GRB
990123. It is possible that another mechanism may be operating
simultaneously which can modify the flux ratio. For example,
down-Comptonization of gamma-ray photons in intervening electron
clouds may be able to reduce the flux ratio substantially if some
gamma-ray photons are converted into optical photons which are also
delayed by several seconds with respect to the gamma-ray emission,
as proposed independently by Zheng et al. (2006). In this model,
different covering factors of high density region around the central
engine could lead to different optical to gamma-ray flux ratios.

\begin{table}
\begin{minipage}[t]{\columnwidth}
\caption{Results of $\chi^2$ tests for different optical lag models}
\label{catalog} \centering
\renewcommand{\footnoterule}{}  
\begin{tabular}{ccccccc}
\hline \hline model \footnote{Refer to text for details.}& $\chi^2$
& $\nu$ & $P(>\chi^2)$ \footnote{The probability that a single
observation from a $\chi^2$ distribution with  degrees of freedom
$\nu$ will be larger than the $\chi^2$ values in column (2).}  &
Lag$_{99}$ \footnote{Time lag for GRB 990123.} (s) & Lag$_{04}$ \footnote{Time lag for GRB 041219a.} (s)\\
\hline
1 & 7.22 & 4 & 0.12 & -- & 0 \\ 
2 & 5150 & 6 & $<10^{-20}$ & 0&0 \\
3 & 6.36 & 4 & 0.17 & 12.2 & 3.0 \\
4 & 7.17 & 5 & 0.21 & 12.2 &5.2 \\
\hline
\end{tabular}
\end{minipage}
\end{table}

\begin{table}
\begin{minipage}[t]{\columnwidth}
\caption{Parameters of GRB 990123 and GRB 041219a based on synchrotron cooling model}
\label{catalog} \centering
\renewcommand{\footnoterule}{}  
\begin{tabular}{lccc}
\hline \hline
Parameter & GRB 990123 & GRB 041219a & Ref.
\footnote{References: 1. Andersen et al. 1999; 2. Kulkarni et al. 1999;
3. Barkov \& Bisnovatyi-Kogan 2005; 4. Barthelmy et al. 2004; 5. Akerlof et al. 1999;
6. Vestrand et al. 2004; 7. Liang et al. 1999; 8. Li et al. 2003; 9. Li et al. 2004.} \\
\hline
Obervation &  &  & \\
Redshift & 1.6 & 0.1  & 1, 2, 3 \\
Time lag (s) & 12.2 & 3.0 &\\
$L_{peak, \gamma}$ (erg/s) & $1.5\times 10^{53}$ & $1.2\times 10^{52}$ & 2, 4\\
$L_{peak, opt}$ (erg/s) & $5\times 10^{49}$ & $1.5\times 10^{47}$ & 5, 6\\
$\nu_{100\ keV}$ (Hz) & $2.4\times10^{19}$ & $2.4\times10^{19}$  &\\
$\nu_{5000 \AA}$ (Hz) & $6\times10^{14}$ & $6\times10^{14}$  &\\
\hline
Assumption &  &  & \\
Bulk Lorentz factor & 300 & 300  & 7 \\
Radius (cm) & $2\times10^{16}$ & $2\times10^{16}$ & 7, 8\\
\hline
Derived parameters &  &  & \\
B (Gs) & 28 & 53 &  \\
$\gamma_\gamma$ & $6.1\times10^{4}$ & $2.9\times10^{4}$  &  \\
$\gamma_{opt}$  & 310  & 140 &  \\
$N_e$ \footnote{Calculated from $L_{peak, \gamma}$.}  &  $3.6\times10^{51}$ & $6.3\times10^{49}$  &  \\
$\nu_{a, \gamma}$ (Hz) &  $6.5\times10^{14}$ & $8.2\times10^{14}$  &  9 \\
$\nu_{a, opt}$ (Hz) &  $6.6\times10^{13}$ & $3.2\times10^{13}$  & 9 \\
$\Sigma$ (cm$^{-2}$) &  $2.6\times10^{23}$ & $4.5\times10^{21}$  &  \\
\hline

\end{tabular}
\end{minipage}
\end{table}

\begin{figure}
\epsfxsize=\hsize\epsfbox{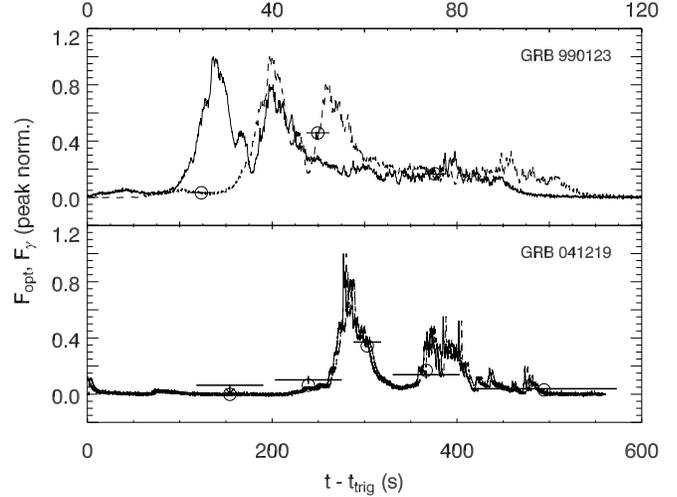} \caption{Observed and simulated
optical and $\gamma$-ray light curves. The upper panel is for GRB
990123 with time indicated in the top axis, and the lower panel is
for GRB 041219a with time indicated in the bottom axis. The optical
flux measured by Akerlof et al. (GRB 990123) or Vestrand et al. (GRB
041219a) are indicated by crosses, where error bars for detections
are given as $1\sigma$ values and non-detections are plotted as
$2\sigma$ upper limits. The corresponding best-fitting simulated
values for model 3 described in the paper are plotted as circles.
The $\gamma$-ray light curves measured by the BASTE (GRB 990123) or
Swift BAT (GRB 041219a) are plotted as solid lines. The best-fitting
simulated optical light curves for model 3 are plotted as dashed
lines.}
\end{figure}

\begin{figure}
\epsfxsize=\hsize\epsfbox{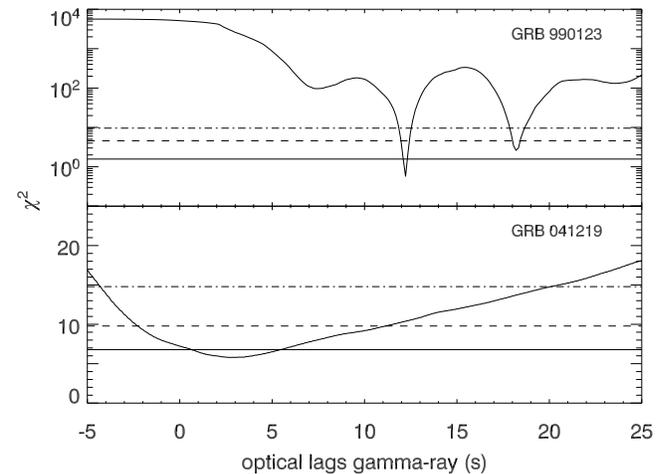} \caption{$\chi^2$ value of simulated optical flux as a
function time lag for model 3 described in the paper. The upper panel is for GRB 990123
and the lower panel is for GRB 041219a. The solid, dashed and dash-dotted horizontal
lines indicate $1\sigma$, $2\sigma$ and $3\sigma$ errors, respectively.}
\end{figure}

\begin{figure}
\epsfxsize=\hsize\epsfbox{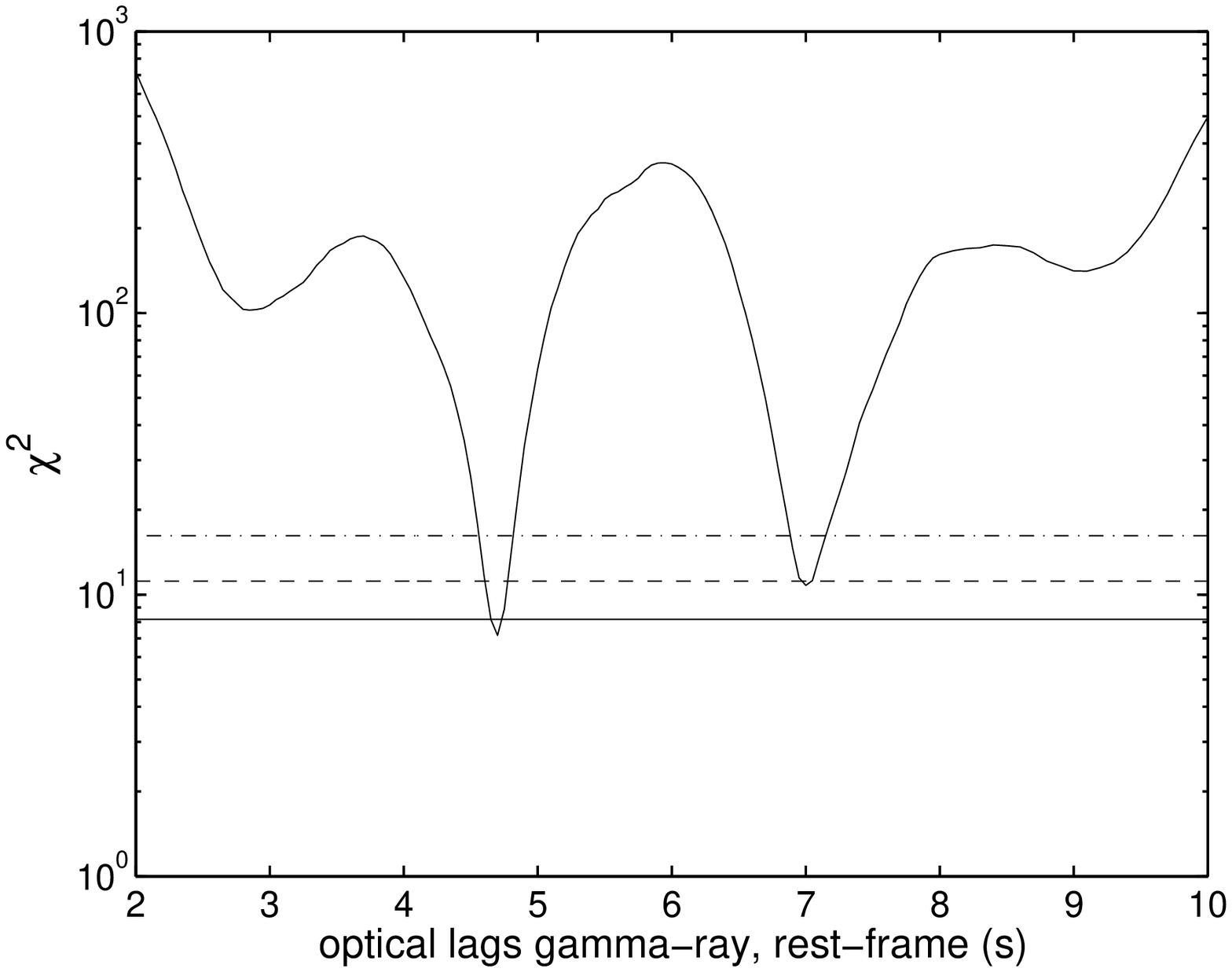} \caption{$\chi^2$ value of simulated optical flux as
a function time lag for both GRB 990123 and GRB 041219a in their host galaxy
rest-frames, for model 4 described in the paper. The solid, dashed and dash-dotted
horizontal lines indicate $1\sigma$, $2\sigma$ and $3\sigma$ errors, respectively.}
\end{figure}

\begin{acknowledgements}
We thank the anonymous referee for helpful comments that improved
the manuscript. SNZ acknowledges partial funding support by the
Ministry of Education of China, Directional Research Project of the
Chinese Academy of Sciences and by the National Natural Science
Foundation of China under project no. 10233010 and 10521001.

\end{acknowledgements}

\end{document}